# A Fluids Experiment for Remote Learners to Test the Unsteady Bernoulli Equation Using a Burette

**Matthew J. Traum**
University of Florida
Gainesville, FL, USA

**Luis Enrique Mendoza Zambrano**
University of Florida
Gainesville, FL, USA

## ABSTRACT

*The COVID-19 pandemic illuminated the critical need for flexible mechanical engineering laboratories simultaneously deployable in multiple modalities: face-to-face, hybrid, and remote. A key element in the lesson portfolio of a forward-looking engineering instructor is economical, hands-on, accessible, "turn-key" lab activities; kits that can be deployed both in brick-and-mortar teaching labs and mailed home to remote learners. The Energy Engineering Laboratory Module (EELM™) pedagogy, described elsewhere, provides an underpinning theoretical framework and examples to achieve these features. In addition, instructional lab kits must demonstrate foundational engineering phenomena while maintaining measurement accuracy and fidelity at reasonable cost. In the energy-thermal-fluid sciences, achieving these conditions presents challenges as kits require energy and matter transport and conversion in real time at scales large enough to reveal measurable phenomena but not so large as to become hazardous to users. This paper presents theoretical underpinning and experimental verification of a fluid mechanics lab experiment appropriate for undergraduate engineering students that 1) meets all the above-described criteria, 2) costs less than $30 in materials, and 3) can be easily mailed to remote learners.*

## NOMENCLATURE

Hereunder, the required nomenclature used to derive the theoretical findings for the control volume analysis is presented.

| | |
|---|---|
| $A$ | cross-sectional area |
| $A_1$ | cross-sectional area of burette water column |
| $A_2$ | cross-sectional area at stopcock valve |
| $c$ | speed of sound in fluid |
| $d_1$ | diameter of burette water column |
| $d_2$ | diameter at stopcock valve |
| $g$ | acceleration due to gravity |
| $K$ | bulk modulus of elasticity of fluid |
| $L$ | water column height |
| $L_o$ | initial water column height |
| $P$ | pressure |
| $t_{drain}$ | drain time associated to inertial motion |
| $t_{est}$ | drain time associated to impulsive motion |
| $t_{info}$ | time of information propagation in the water column |
| $t_{total}$ | total drain time |
| $v$ | flow velocity |
| $z$ | elevation of fluid with respect to reference datum |
| $z_1$ | initial water elevation with respect to reference datum |
| $z_2$ | final water elevation with respect to reference datum |
| $\rho$ | water density |

## 1. INTRODUCTION

Bernoulli's equation is often taught as the interchangeability between kinetic, potential, and pressure energy in moving fluid under energy conservation assumptions. However, the additional unsteady term representing convective energy to either accelerate or decelerate fluid streamlines is almost always neglected owing to assumption of steady state flow. Nonetheless, the unsteady Bernoulli's equation models numerous fluids phenomena induced by impulsive action including water hammer, manometer oscillations, Coriolis flow metering, and the first moments of tank draining. Thus, its coverage in undergraduate fluids courses is beneficial.

At the core of the experiment reported in this paper is a 50 mL acrylic burette with manual stopcock to regulate flow. This robust instrument is commonly used in high school and lower-division college chemistry laboratories for titration experiments. For this fluids experiment, learners fill the burette with water, determine the static water column height, and then fully open the stopcock to facilitate draining. The burette's accurate scale and ability to drain to a precise, reproduceable volume enables transient experiments with repeatable initial conditions.

Manually clocking the burette draining process enough times ($N \geq 12$) to conduct statistical uncertainty analysis establishes experimental drain time and uncertainty in this value. Comparing experimental results to theoretical drain time derived from Bernoulli's equation reveals need to consider both steady and unsteady expressions to predict drain time correctly. This lab experiment demonstrates the utility of simple and flexible activities to illuminate important energy-thermal-fluid science phenomena for students in multiple simultaneous learning modalities, especially remote and hybrid where students use mailed kits to facilitate labs with hands-on learning.

## 2. BACKGROUND

An incredible tradition exists in college chemistry and physics courses for using simple glassware and available lab components to demonstrate engineering phenomena. For example, Muiño and Hodgson used an Erlenmeyer flask to create Hero's Engine [1]; Jackson and Laws used syringes to make thermometers, hydraulic pumps, and heat engines [2]; and Steffensena combined a precision bore Pyrex tube with an inexpensive pressure transducer to perform Rüchardt's gas specific heat ratio measurement [3].

The burette-based experiment described here follows in spirit the works of these teacher/scholars. It is an elegant but tractable experiment that uses simple and readily available laboratory components, and it enables students to explore relevant engineering phenomena themselves. However, to be effective and adopted by college and high school instructors, such experiments must be hands-on, accessible, student-centered, economical, and "turn-key". The project hardware must be affordable for an institution with limited resources, and the experiments must be buildable and operable by students without situated knowledge or access to specialized tools. The Energy Engineering Laboratory Module (EELM™) pedagogy, which is described elsewhere [4-5], provides underpinning theoretical pedagogical framework. Examples that achieve these features at the college level include 1) a series of building energy audit exercises that harvests existing buildings as living laboratories suitable for quantitative evaluation using an inexpensive audit tool kit [6] and a small, inexpensive inverted downdraft wood gasifier for processing pine chips into syngas using metal vacuum-flask-style thermos bottle [7]. Examples at the high school level include 1) Design / Build / Fly / Analyze / Redesign / Build / Fly rocket curricula [8-9] and use of standard benchtop physics laboratory equipment to introduce hands-on machining fabrication techniques in science classes [10].

In the energy-thermal-fluid sciences fields, achieving EELM™ conditions presents challenges as the experiments require energy and matter transport and conversion in real time at scales large enough to reveal measurable phenomena but not so large as to become hazardous to users. Thus, additional benefit is obtained by structing these experiments in kit form where they are both contained in size but also capable of being used in multiple learning modalities (in-person, remote, and hybrid) by students without need for instructor or technician intervention.

While no fully online ABET-accredited mechanical engineering bachelor's degree programs yet exists [11], examples from the mechanical engineering education literature show that online laboratory instruction matches or exceeds traditional face-to-face education with benefits including 1) increased accessibility, 2) increased retention and completion, 3) increased flexibility, and 4) increased interaction and engagement [12].

In fact, the literature shows that many students favor remote lab experiences over on-site laboratories [13-14]. For example, Corter and colleagues explored student achievement of learning objectives using cantilever beam experiments where content was delivered in three different environments: 1) brick-and-mortar labs, 2) lab kits at home, and 3) remote labs via computer simulation. These researchers found that at home, hands-on experiments were at least as effective as traditional brick-and-mortar labs for student learning. Moreover, student survey responses favored the remote lab experiences and remote students performed better in outcome achievement [15-16]. Investigations of kit-based lab instruction in other engineering disciplines agree with Corter's conclusion that remote hands-on experiments are better for student learning than traditional labs. Students gain experience building and running their own experiments, and they can explore interesting or unexpected observations at will without the time constraint of a classroom lab schedule.

Further benefits include ease of tailoring experiments to meet student needs as well as flexibility of laboratories to be shared between universities [17]. In a present example, instructors at the University of Florida and Aswan University in Egypt are organizing a lab-based virtual exchange program where students from both institutions complete lab experiments remotely and collaboratively for fluids lab courses co-taught simultaneously at both institutions. This exchange facilitates both technical and culture discourse for large engineering student populations not possible in brick-and-mortar teaching environments [18]. A complete fluid mechanics lab kit has been posited [19] and experiments successfully demonstrated including a circular hydraulic jump [20], a benchtop fan curve performance measurement apparatus [21], and an internal pipe flow velocity interrogator [22-23]. Individual experiments for a future thermodynamics kit are also in development, including a desktop-scale Tesla turbine [24] tied to a Prony Brake dynamometer capable of demonstrating aerospace turbomachinery operation [25].

When used in an in-person instructional environment, kits enable a single space to be used for multiple different classes in parallel to make more efficient use of limited space. This approach was used at Jacksonville University to teach electronics engineering lab, mechanics of materials lab, and product develop courses in the same physical laboratory space in one semester [26-27]. If an incident drives instruction all-online again, students using the kits in a brick-and-mortar setting can simply take the kits home and continue the lab class remotely without disruption.

### 3. EXPERIMENTAL SEP-UP

The experiment is conducted using a 50 mL capacity EISCO CH0233B Acrylic Burette with 0.1 mL graduations and a PTFE stopcock ($19.75 on Amazon.com at time of writing) mounted vertically to a sturdy table using a mini jeweler clamp ($5.99 on Amazon.com at time of writing) as shown in Figure 1.

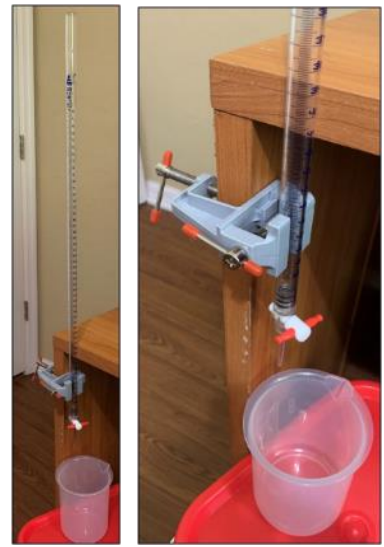

**FIGURE 1:** (LEFT) COMPLETE BURETTE SETUP CLAMPED TO THE LEG OF A DESK AT HOME. (RIGHT) CLOSE-UP VIEW OF THE BURETTE STOPCOCK FROM WHICH FLUID DRAINS INTO A CATCHMENT BEAKER.

The burette's verticality was verified within ±0.1° of 0° with the free Bubble Level iOS app available for smart phones. Similarly, a free smartphone stopwatch was used to directly measure the experimental drain time, which is later compared to theoretical results. Each lab kit will be included as an additional course fee to the students' course cost.

## 4. THEORETICAL DEVELOPMENT

Derivation of Bernoulli's equation under the assumption of fluid flow without energy loss is demonstrated elsewhere from both top-down application of Reynold's Transport Theorem [28] or bottom-up application of Euler's equations [29].

An appropriate burette engineering model is shown in Figure 2. The system is a tall right cylindrical column of liquid water in a container with a circular hole in the bottom much smaller than the container diameter. The drain is initially blocked by the stopcock valve that when opened instantaneously allows the water to flow under gravity out of the column base.

In general, with the stopcock open, for incompressible fluid along a streamline from Station 1 to Station 2 in a generic flow field gives the following relationship:

$$P_1 + \frac{\rho v_1^2}{2} + \rho g z_1 = P_2 + \frac{\rho v_2^2}{2} + \rho g z_2 + \int_1^2 \rho \frac{\partial v}{\partial t} ds \quad (1)$$

Further assuming that flow is in steady state gives the familiar Bernoulli's equation form, dropping the final unsteady term:

$$P_1 + \frac{\rho v_1^2}{2} + \rho g z_1 = P_2 + \frac{\rho v_2^2}{2} + \rho g z_2 \quad (2)$$

Equation 2 models inviscid flow unchanging with time, but it does not account for the inertial effects in fluids whose motion is impulsively initiated when the stopcock is opened. For those situations, the final term in Equation 1 cannot be neglected, and its presence sometimes creates unanticipated behaviors of the kind at the focus of this experiment.

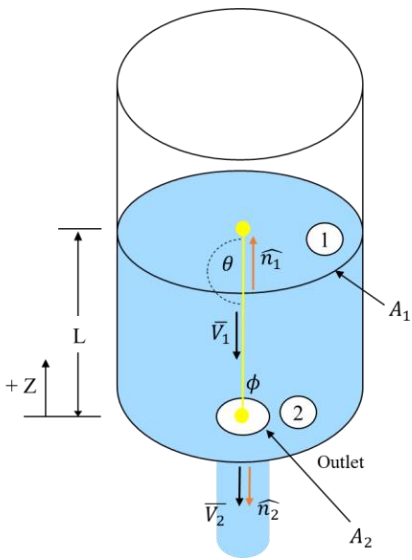

**FIGURE 2:** A SCHEMATIC REPRESENTATION OF THE EXPERIMENTAL BURETTE COLUMN FILLED WITH LIQUID WATER SHOWING THE ANTICIPATED STREAMLINE FLOW FROM STATES 1 TO STATE 2 USED IN EQUATION 1.

The experiment interrogates two interconnected unsteady flow phenomena. First is the time required after the valve is opened for the fluid to fully begin moving (Equation 1). Second is application of Bernoulli's equation in steady state (Equation 2) to describe the fluid motion once the impulsive event that initiated motion has passed. Since the water column height inducing hydrostatic pressure on the drain hole is becoming smaller with time, as this height/pressure falls so does the velocity of flow out of the hole; as the water column becomes shorter it drains more slowly. Stitching together these two phenomena gives an engineering estimate for column drain time.

There is also a third time scale: the rate of information propagation in the water column. From the moment the stopcock valve is opened, some time elapses before the fluid on the water column's free surface is "aware" the valve has moved,

$$t_{info} = \frac{L_o}{c} \quad (3)$$

This information propagates as a pressure wave up the column at the speed of sound in fluid.

$$c = \left(\frac{K}{\rho}\right)^{\frac{1}{2}} \quad (4)$$

However, the propagation time is ~3.6 x 10$^{-4}$ seconds, much shorter than times associated with the other flow phenomena in this experiment. It is therefore ignored in calculations.

### 4.1 Impulsive Motion Study

In the first draining phase, the valve at the burette column bottom is opened instantaneously, and the water experiences an impulsive change that induces flow. The duration it takes the water column to fully react to the impulsive valve opening is found by applying Equation 1 along a streamline in the water column of Figure 2 from Station 1, the free surface, to Station 2, the location of the opening valve.

With the valve opened, the pressure on the free surface is atmospheric as is the jet issuing from the valve. As both are atmospheric pressure, they cancel in Equation 1. Moreover, the coordinate system in Figure 2 establishes $z = 0$ at Station 2 and $z = L_o$ at Station 1. Applying this information, Equation 1 becomes:

$$\frac{\rho v_1^2}{2} + \rho g L_o = \frac{\rho v_2^2}{2} + \int_1^2 \rho \frac{\partial v}{\partial t} ds \quad (5)$$

The final form of the equation is obtained by substituting a relationship between $v_1$ and $v_2$ found by applying Mass Conservation to the control volume (Figure 2) and remembering that the fluid is incompressible. Here, $A_1$ is the cross-sectional area of the burette water column and $A_2$ is the cross-sectional area at the stopcock valve.

$$v_1 = \frac{A_1}{A_2} v_2 \tag{6}$$

$$\int_1^2 \rho \frac{\partial v_2}{\partial t} ds + \frac{\rho}{2}\left(1 - \left(\frac{A_2}{A_1}\right)^2\right) v_2^2 - \rho g L_o = 0 \tag{7}$$

To carry out the integration of the first term in Equation 7, two system attributes are recognized: a) the distance $ds$ is an infinitesimally small step along the streamline connecting Stations 1 and 2 with initial height $L_o$ and b) in this initial period, velocity along the streamline changes with time but not position. In other words, the free surface of the water column does not move the instant the valve is opened, which establishes the boundary condition $v_1(t = 0) = v_2 = 0$. Thus, the first term in Equation 7 can be integrated, giving after some algebra:

$$\frac{\partial v_2}{\partial t} + \frac{1}{2L_o}\left(1 - \left(\frac{A_2}{A_1}\right)^2\right) v_2^2 - g = 0 \tag{8}$$

Equation 8 is a first order differential equation of the following form:

$$\frac{\partial v}{\partial t} + Mv^2 - N = 0 \tag{9}$$

which has solutions of the form:

$$v(t) = \frac{\sqrt{N} \tanh\left(c_1 \sqrt{MN} + t\sqrt{MN}\right)}{\sqrt{M}} \tag{10}$$

After solving the integration constant at the initial boundary condition, the final equation becomes:

$$v(t) = \frac{\sqrt{2L_o g} \tanh\left(t \sqrt{\frac{g}{2L_o}\left[1 - \left(\frac{A_2}{A_1}\right)^2\right]}\right)}{\sqrt{\left[1 - \left(\frac{A_2}{A_1}\right)^2\right]}} \tag{11}$$

To estimate the duration needed for the water column to fully react to the impulsive change caused by opening the stopcock, the steady state velocity, $v(t \to \infty)$ is needed. Since the hyperbolic tangent in Equation 11 approaches 1 asymptotically for $t \to \infty$, Equation 11 simplifies to the steady state velocity:

$$v(t) = \sqrt{\frac{2L_o g}{\left[1 - \left(\frac{A_2}{A_1}\right)^2\right]}} \tag{12}$$

However, a pragmatic problem remains. The asymptotic nature of the hyperbolic tangent in Equation 11 means it takes infinite time for this system to reach the steady state velocity in Equation 12. An engineering convention, "Establishment Time" ($t_{est}$), is set by stating a final velocity at 99% of the steady state value is close enough for engineering purposes.

$$0.99 \approx \tanh\left(t_{est} \sqrt{\frac{g}{2L_o}\left[1 - \left(\frac{A_2}{A_1}\right)^2\right]}\right) \tag{13}$$

Rearranging Equation 13 to solve for $t_{est}$ gives:

$$t_{est} \approx \frac{\operatorname{arctanh}(0.99)}{\sqrt{\frac{g}{2L_o}\left[1 - \left(\frac{A_2}{A_1}\right)^2\right]}} \tag{14}$$

Geometrically, both the water column and drain hole in this experiment are circular cross sections measurable by caliper, and the $t_{est}$ becomes:

$$t_{est} \approx \frac{\operatorname{arctanh}(0.99)}{\sqrt{\frac{g}{2L_o}\left[1 - \left(\frac{d_2}{d_1}\right)^4\right]}} \tag{15}$$

As an additional simplifying estimate is $d_1 \gg d_2$, then $\left(\frac{d_2}{d_1}\right)^4 \ll 1$ and the diameter ratio in the denominator can be neglected to leave:

$$t_{est} \approx \frac{\operatorname{arctanh}(0.99)\sqrt{2L_o}}{\sqrt{g}} \tag{16}$$

This simplified expressional form reveals the underlying physics that $t_{est}$ for the water column to approach steady state velocity in response to an impulse event is proportional to $\sqrt{L_o}$. Intuitively, the taller the water column, the longer it takes the fluid to overcome its inertia.

**4.2 Inertial Motion Study**

Bernoulli's equation is also useful for estimating burette water column drain time after the Establishment Time has elapsed from the initial opening impulse. This analysis follows from Equation 2, the steady Bernoulli's equation, but it unfolds differently than the impulsive unsteady analysis because focus is in $v_1$, the free surface fluid velocity.

When the stopcock valve is opened instantaneously, the pressure on the free surface matches the pressure of the water just coming out of the valve. Moreover, the coordinate system establishes z = 0 at Station 2 and z = L at Station 1. Recasting Equation 2, it becomes:

$$2gL = v_2^2 - v_1^2 \tag{17}$$

Another key difference between this analysis and the previous unsteady one is that the fluid column height, $L$, is now variable because the column is draining, not just reacting to an impulse boundary change. Here, water column height changes as a function of time: $v_1 = -\frac{dL}{dt}$. Applying the mass conservation expression of Equation 6 to Equation 17 gives:

$$2gL(t) = \left[\left(\frac{A_1}{A_2}\right)^2 - 1^2\right]\left(-\frac{dL}{dt}\right)^2 \tag{18}$$

This expression is a differential equation with the following solution, again recognizing that both the water column and drain hole in the experiment are circular cross sections.

$$L(t) = \left( [L_o]^{\frac{1}{2}} - \frac{t}{2}\sqrt{\frac{2g}{\left[\left(\frac{d_1}{d_2}\right)^4 - 1\right]}} \right)^2 \quad (19)$$

Equation 19 gives the instantaneous water column height, $L(t)$, for any elapsed time, $t$, after the Establishment Time of the initial impulsive unsteady flow period has elapsed. Drain time for this portion of the process is found by setting the final height of the water column to zero, its height when the water has fully drained, and solving for $t_{drain}$.

$$t_{drain} = \sqrt{\frac{2L_o\left[\left(\frac{d_1}{d_2}\right)^4 - 1\right]}{g}} \quad (20)$$

An interesting difference between the inertial and impulsive analyses is for the impulsive component the burette geometry where $d_1 \gg d_2$ diminishes effects of the geometry whereas for the inertial component the opposite occurs. The larger the brunette diameter, $d_1$, gets relative to the stopcock hole, $d_2$, the longer is this later drain time. The ratio of drain hole to burette dimeters matters for inertial flow but not impulsive flow.

### 4.3 Complete Drain Time Equation
The ultimate goal is developing an engineering estimate for total burette drain time by summing the elapsed time of the impulsive plus inertial flow for this system.

$$t_{total} = t_{est} + t_{drain} \quad (21a)$$

$$t_{total} = \frac{\operatorname{arctanh}(0.99)\sqrt{2L_o}}{\sqrt{g}\left[1 - \left(\frac{d_2}{d_1}\right)^4\right]} + \sqrt{\frac{2L_o\left[\left(\frac{d_1}{d_2}\right)^4 - 1\right]}{g}} \quad (21b)$$

This expression will be compared against experimental drain time measured for the burette to validate this modeling approach.

### 5. METHODS
The burette mounted as shown in Figure 1 was overfilled with room temperature water and the stopcock valve closed. The valve was then opened, and water was bled out until the fluid meniscus was level with the 0 mL volume mark, as shown in Figure 3. The burette is sized so each 1 mL mark corresponds to 1 cm of water height. There are volume demarcations at 0.1 mL intervals, which allow the water to be set repeatably at $\pm 0.05$ mL volumes corresponding to $\pm 0.05$ cm uncertainty in initial height. The burette markings increase from 0 mL to 50 mL as the fluid height drops indicating volume of liquid delivered, and there is a dead volume at the burette's bottom. Measured by a caliper ($\pm 0.1$ cm accuracy), the stopcock valve hole begins 3.9 cm below the 50 mL burette mark. Thus, the initial height of water in the burette is taken as $53.9 \pm 0.1$ cm. The burette inner diameter and stopcock valve inner diameter were measured by caliper ($\pm 0.002$ cm accuracy) at $1.100 \pm 0.002$ cm and $0.130 \pm 0.002$ cm, respectively. The drain experiment was repeated 12 times; enough to collect statistically meaningful average and standard deviation for drain time.

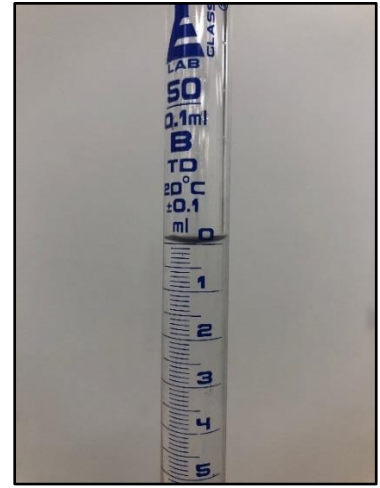

**FIGURE 3:** FOR EACH OF 12 EXPERIMENTAL RUNS THE BURETTE WAS FILLED TO $50.0 \pm 0.1$ mL TOTAL VOLUME WITH ROOM TEMPERATURE WATER.

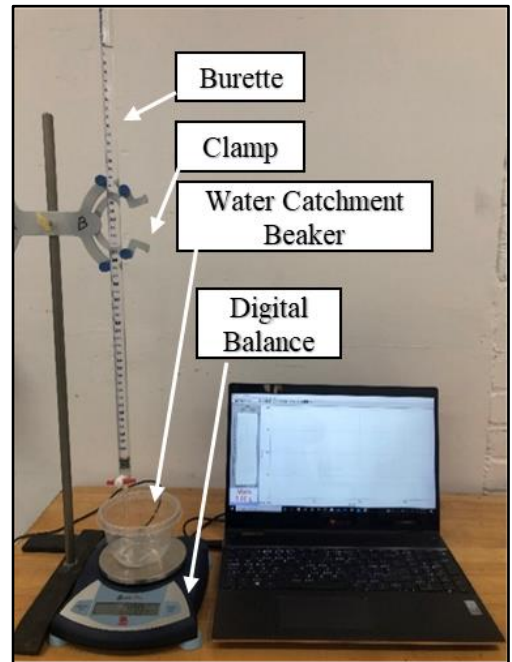

**FIGURE 4:** SET-UP INCORPORATING A DIGITAL BALANCE FOR DATA COLLECTION TO GENERATE VISUAL REPRESENTATIONS OF TYPICAL DRAINING EXPERIMENTS.

Although a balance would not be used in the students' laboratory experiment, to generate visual representations for typical draining experiments, an Ohaus SP602 digital balance was placed under the water catchment beaker as shown in Figure

4. It was connected to a data acquisition computer running Vernier Logger Pro 3 software set to collect accumulated mass readings once per second. To calibrate and validate conditions and equipment specific to the lab, the density of water used in the experiment was measured. The Ohaus SP602 balance determined the mass of a corresponding measured volume of water drained from the burette. The room temperature and local atmospheric pressure were 23.8°C and 101.1 kPa, respectively during this test, and the accepted water density value under these conditions is 0.99865 g/mL. The experimental density measurement was 1.014 g/mL, an error of 1.5% from the accepted value.

For each unique experiment in which the stopwatch was used, the burette was filled with water and bled to the 50 mL mark. The stopcock was then closed. Next, the experimenter simultaneously turned the stopcock to its fully open position and started a stopwatch. The watch was stopped coincident with the last water leaving the burette. To generate a measurement uncertainty at the 95% confidence level, plus or minus two standard deviations for the drain time data are reported. The resulting experimental drain time for the 50 mL burette was 24.2 ± 0.4 seconds. Figure 5 shows a representative sample plot from one of the 12 burette drain time runs.

## 6. RESULTS

Additionally, as seen in Figure 5, the mass flow rate associated with the fluid draining out of the burette changes with the water column's height.

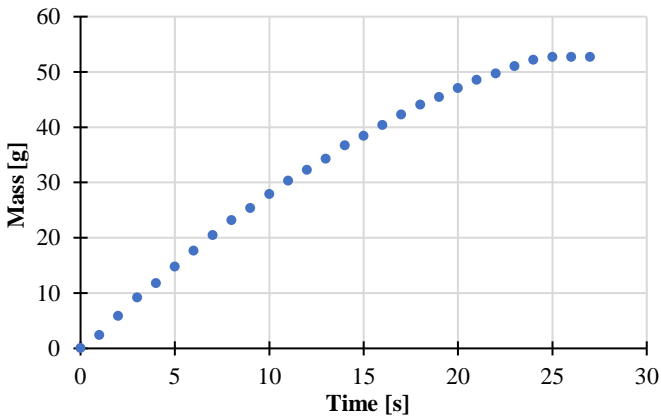

**FIGURE 5:** DRAIN TIME VERSUS CATCHMENT BEAKER MASS ACCUMULATION FOR THE EXPERIMENTAL DRAINING OF A 50 ML WATER-FILLED BURETTE.

This relationship is appreciated by noticing that the slope of the curve in Figure 5 is not constant. In addition, during the initial interval, the average rate of change between the first two data points is in fact smaller than in the next pair of consecutive measurements. This feature illustrates that there is indeed a time required after the stopcock valve is opened for the fluid to fully begin moving. Therefore, the previously described motion portrayed by the unsteady Bernoulli equation in Equation 15 is a real phenomenon that requires inclusion in the analysis for accurate drain time prediction. From Figure 5, the drain time is known to be between 24 and 25 seconds since after the 25-second mark the accumulated mass remains constant, indicating the burette has fully drained. Better experiential drain time precision could not be achieved because the Ohaus digital balance cannot sample faster than one measurement per second based on its device interface. However, it is possible to validate the experimental result obtained by employing a stopwatch since this value is within the interval of the solution giving by the mass-time measurements.

### 6.1 Experiment / Model Comparison Results

Inserting caliper-measured numbers into Equations 15 and 20 respectively gives $t_{est} = 0.9\ s$ and $t_{drain} = 23.7\ s$ with the sum being $t_{total} = 24.6\ s$ as the predicted drain time. The acceleration due to gravity was taken as 9.806 m/s$^2$ for these calculations. This result matches the experimental measurement to within experimental uncertainty.

Furthermore, the uncertainty in the theoretical drain time was computed based on the error in the caliper measurements and initial height of the water column. The Root Sum Squared (RSS) method was used to propagate error from measurements. Consequently, the predicted drain time was found as 24.6 ± 0.7 seconds. Figure 6 compares predicted versus experimental drain times. It is important to highlight that by just considering the inertial component of Bernoulli's equation, the predicted drain time still falls within the experimental results as its uncertainty alone is greater than 0.7 s. Moreover, the impulsive component has negligible uncertainty.

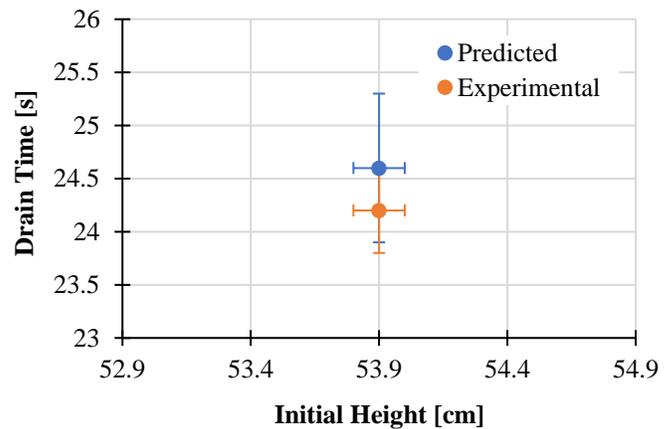

**FIGURE 6:** COMPARISON BETWEEN EXPERIMENTAL AND PREDICTED DRAIN TIME HIGHLIGHTING ERROR BARS.

## 7. DISCUSSION

The theoretical model, which contains both impulsive (unsteady) and inertial (steady) components derived using Bernoulli's equation agrees with measured results within the uncertainty of the experiment. Although it represents only about 4% of the duration of the total burette drain time, the unsteady analysis component is an important element to correctly predicting drain time as it explains the slope discontinuity in Figure 5 between t = 0 s and t = 1 s. Without considering the unsteady component, the steady model alone could underpredict the actual drain time and would not agree within experimental uncertainty of the measured result. The smallest experimental time based on the uncertainty in the stopwatch-measurements corresponds to 23.8 s while the predicted time associated to the inertial motion of the fluid is $t_{drain} = 23.7 \pm 0.7\ s$. Therefore, an improvement in the accuracy of measurements is needed in calculating the theoretical drain time to reduce uncertainty and isolate the unsteady term in the analysis.

The uncertainty in total theoretical time as well as the isolated time related to inertial motion is mostly dependent on the experimenter's ability to accurately measure the diameter of the stopcock valve. A caliper or alternative length measuring instrument with an accuracy better than 0.001 mm is needed. At this measurement resolution, uncertainty in the theoretical drain time diminishes to a point where contribution of the impulsive unsteady Bernoulli process can be resolved. Consequently, results with more accurate stopcock valve opening diameter measurements part of future work. In addition, comparative methods developed to highlight the importance of the unsteady term in Bernoulli's equation for burette draining will be discussed in future papers.

## 8. CONCLUSION

Theoretical underpinning and experimental verification of a fluid mechanics lab experiment illustrating application of the unsteady Bernoulli equation and appropriate for undergraduate engineering (and even high school physics) teaching labs were presented. The essential experimental components cost about $25 (at time of writing), and the experiment can be successfully converted into a kit for both in-person and remote teaching laboratory settings with emphasis on students being able to safely complete the experiment from home.

The theoretical model, which contains both impulsive (unsteady) and inertial (steady) components derived using Bernoulli's equation predicts drain time of 24.6 ± 0.7 seconds, which agrees with the experiment, which measured drain time at 24.2 ± 0.4 seconds.

Although it represents only about 4% of the total burette drain time duration, the unsteady analysis component seems to be essential for correctly predicting drain time and interpreting graphical data collected with a digital data acquisition system. The unsteady component explains measured discontinuity in the experimental drained mass versus time slope at 0 ≤ t < 1 seconds. Without considering the unsteady component, the steady model alone underpredicts the actual drain time. This outcome illustrates the importance of including the unsteady Bernoulli equation in analyses in which fluids reacts to impulsive changes in boundary conditions.